\documentclass[graybox,vecphys]{svmult}

\usepackage{makeidx}         
\usepackage{graphicx}        
\usepackage{epsfig}                             
        
\usepackage{cite}        
                             
\usepackage{multicol}        
\usepackage[bottom]{footmisc}

\usepackage{newtxtext}       %
\usepackage{newtxmath}       

\usepackage{url} 

\makeindex             

\begin{document}

\newcommand{\jcmindex}[2]{\index{{\bf\large #1}!#2}}
\newcommand{\jcmindext}[3]{\index{{\bf\large #1}!#2!#3}}


\newcommand{\tdt}{t+\Delta t}



\title*{Nonlinearity + Networks: A 2020 Vision}
\author{Mason A. Porter
}

\institute{
Mason A. Porter
\at
Department of Mathematics, University of California, Los Angeles, Los Angeles, California 90095, USA
\email{mason@math.ucla.edu}
}
\maketitle

\vspace{-4 cm}

\abstract{I briefly survey several fascinating topics in networks and nonlinearity. I highlight a few methods and ideas, including several of personal interest, that I anticipate to be especially important during the next several years. These topics include temporal networks (in which the entities and/or their interactions change in time), stochastic and deterministic dynamical processes on networks, adaptive networks (in which a dynamical process on a network is coupled to dynamics of network structure), and network structure and dynamics that include ``higher-order'' interactions (which involve three or more entities in a network). I draw examples from a variety of scenarios, including contagion dynamics, opinion models, waves, and coupled oscillators. 
}



\section{Introduction} \label{sec1}

Network analysis is one of the most exciting areas of applied and industrial mathematics \cite{newman2018book,howison-prsa,mason-ginestra}. It is at the forefront of numerous and diverse applications throughout the sciences, engineering, technology, and the humanities. The study of networks combines tools from numerous areas of mathematics, including graph theory, linear algebra, probability, statistics, optimization, statistical mechanics, scientific computation, and nonlinear dynamics. 

In this chapter, I give a short overview of popular and state-of-the-art topics in network science.
 My discussions of these topics, which I draw preferentially from ones that relate to nonlinear and complex systems, will be terse,
  but I will cite many review articles and highlight specific research papers for those who seek more details.
  This chapter is not a review or even a survey; instead, I give my perspective on the short-term and medium-term future of network analysis in applied mathematics for
  2020 and beyond. 

My presentation proceeds as follows. In Section \ref{sec2}, I review a few basic concepts from network analysis. In Section \ref{sec3}, I discuss the dynamics of networks in the form of time-dependent (``temporal'') networks. In Section \ref{sec4}, I discuss dynamical processes --- both stochastic and deterministic --- on networks. In Section \ref{sec5}, I discuss adaptive networks, in which there is coevolution of network structure and a dynamical process on that structure. In Section \ref{sec6}, I discuss higher-order structures (specifically, hypergraphs and simplicial complexes) that aim to go beyond the standard network paradigm of pairwise connections. I conclude with an outlook in Section \ref{sec7}.


\section{Background on Networks} \label{sec2}

\begin{figure}
\centering
(a) \epsfig{figure=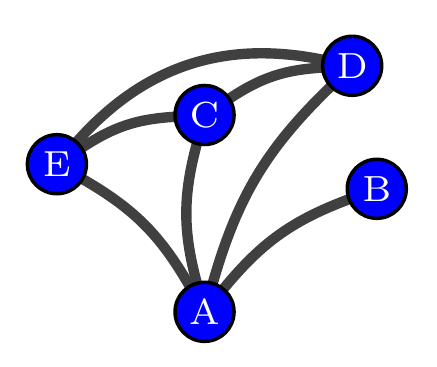,height=2.4cm}
(b) \epsfig{figure=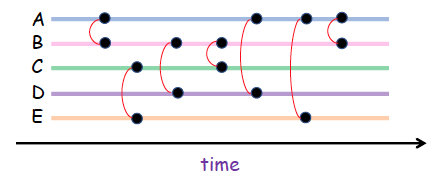,height=2.4cm}\\
\vspace{.2cm}
(c) \epsfig{figure=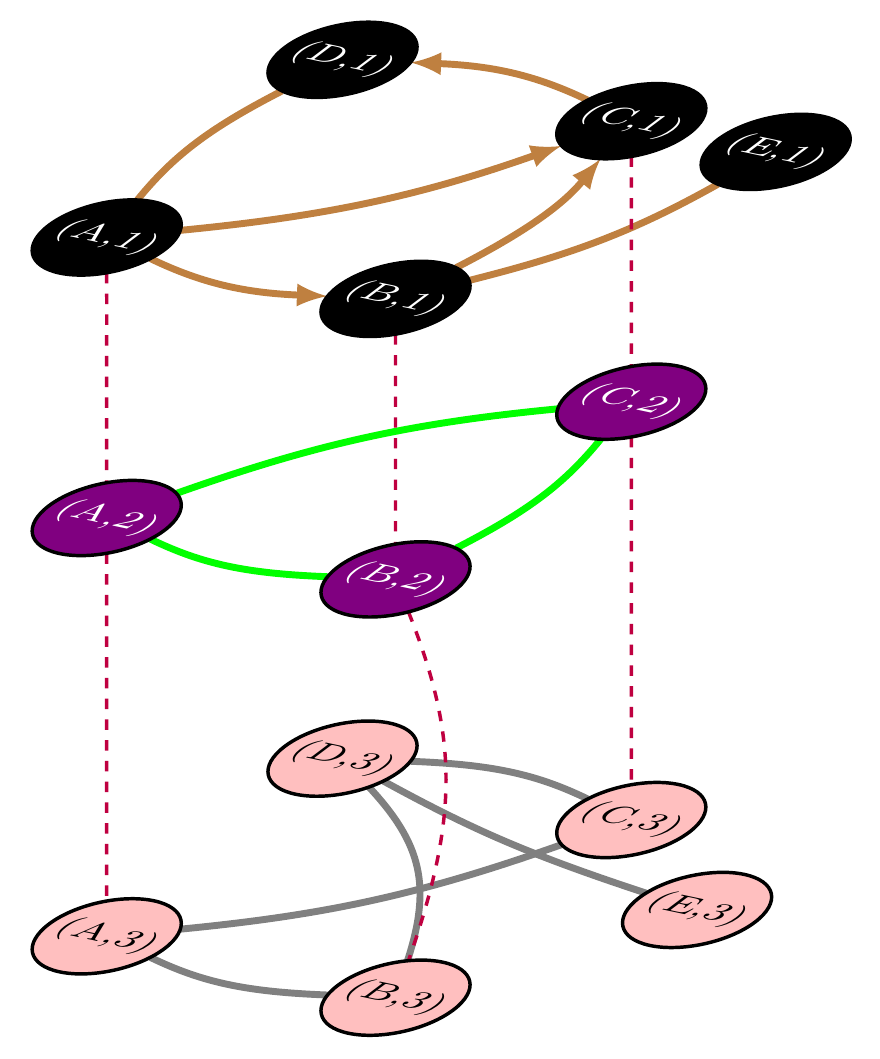,height=4.7cm}
(d) \epsfig{figure=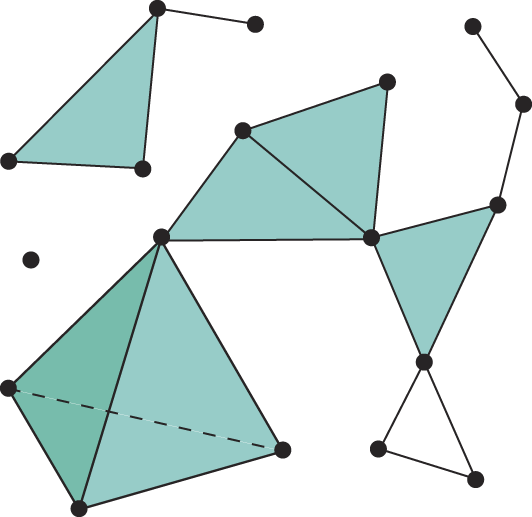,height=4.7cm}
\caption{Several types of network structures: (a) a graph, (b) a temporal network, (c) a multilayer network, and (d) a simplicial complex.
[I drew panels (a) and (c) using {\sc Tikz-network}, which is by J\"urgen Hackl and is available at {\tt https://github.com/hackl/tikz-network}. Panel (b) is inspired by Fig.~1 of \cite{holme2012}. Panel (d), which is in the public domain, was drawn by Wikipedia user Cflm001 and is available at {\tt https://en.wikipedia.org/wiki/Simplicial\_complex}.]
}\label{fig1}
\end{figure}

In its broadest form, a network consists of the connectivity patterns and connection strengths in a complex system of
interacting entities \cite{newman2018book}. The most traditional type of network is a graph $G = (V,E)$ (see Fig.~\ref{fig1}a), where $V$ is a set of ``nodes'' (i.e., ``vertices'') that encode entities and $E \subseteq V \times V$ is a set of ``edges'' (i.e., ``links'' or ``ties'') that
encode the interactions between those entities. However, recent uses of the term ``network'' have focused increasingly on connectivity patterns that are more general than graphs \cite{renaud2019}: a network's nodes and/or edges (or their associated weights) can change in time \cite{holme2012,holme2015} (see Section \ref{sec3}), nodes and edges can include annotations \cite{chodrow2019b}, a network can include multiple types of edges and/or multiple types of nodes \cite{kivela2014,whatis2018}, it can have associated dynamical processes \cite{porter2016} (see Sections \ref{sec3}, \ref{sec4}, and \ref{sec5}), it can include memory \cite{rosvall-memory2014}, connections can occur between an arbitrary number of entities \cite{otter2017,petri-review} (see Section \ref{sec6}), and so on.

Associated with a graph is an adjacency matrix ${\bf A}$ with entries $a_{ij}$. In the simplest scenario, edges either exist or they don't. If edges have directions, $a_{ij} = 1$ when there is an edge from entity $j$ to entity $i$ and $a_{ij}=0$ when there is no such edge. When $a_{ij}=1$, node $i$ is ``adjacent'' to node $j$ (because we can reach $i$ directly from $j$), and the associated edge is ``incident'' from node $j$ and to node $i$. The edge from $j$ to $i$ is an ``out-edge'' of $j$ and an ``in-edge'' of $i$. The number of out-edges of a node is its ``out-degree'', and the number of in-edges of a node is its ``in-degree''. For an undirected network, $a_{ij}=a_{ji}$, and the number of edges that are attached to a node is the node's ``degree''. One can assign weights to edges to represent connections with different strengths (e.g., stronger friendships or larger transportation capacity) by defining a function $w: E \longrightarrow \mathbb{R}$. In many applications, the weights are nonnegative, although several applications \cite{traag2018} (such as in international relations) incorporate positive, negative, and zero weights. In some applications, nodes can also have self-edges and multi-edges. The spectral properties of adjacency (and other) matrices give important information about their associated graphs \cite{newman2018book,piet-book}. For undirected networks, it is common to exploit the beneficent property that all eigenvalues of symmetric matrices are real.

\section{Time-Dependent Networks} \label{sec3}

Traditional studies of networks consider time-independent structures, but most networks evolve in time. For example, social networks of people and animals change based on their interactions, roads are occasionally closed for repairs and new roads are built, and airline routes change with the seasons and over the years. To study such time-dependent structures, one can analyze ``temporal networks''. See \cite{holme2012,holme2015} for reviews and \cite{holme2013,holme2019} for edited collections.

The key idea of a temporal network is that networks change in time, but there are many ways to model such changes, and the time scales of interactions and other changes play a crucial role in the modeling process. There are also other important modeling considerations. To illustrate potential complications, suppose that an edge in a temporal network represents close physical proximity between two people in a short time window (e.g., with a duration of two minutes). It is relevant to consider whether there is an underlying social network (e.g., the friendship network of mathematics Ph.D. students at UCLA) or if the people in the network do not in general have any other relationships with each other (e.g., two people who happen to be visiting a particular museum on the same day). In both scenarios, edges that represent close physical proximity still appear and disappear over time, but indirect connections (i.e., between people who are on the same connected component, but without an edge between them) in a time window may play different roles in the spread of information. Moreover, network structure itself is often influenced by a spreading process or other dynamics, as perhaps one arranges a meeting to discuss a topic (e.g., to give me comments on a draft of this chapter). See my discussion of adaptive networks in Section \ref{sec5}.


\subsection{Discrete Time}

For convenience, most work on temporal networks employs discrete time (see Fig.~\ref{fig1}(b)). Discrete time can arise from the natural discreteness of a setting, discretization of continuous activity over different time windows, data measurement that occurs at discrete times, and so on.


\subsubsection{Multilayer representation of temporal networks} \label{multitemp}

One way to represent a discrete-time (or discretized-time) temporal network is to use the formalism of ``multilayer networks'' \cite{kivela2014,whatis2018}. One can also use multilayer networks to study networks with multiple types of relations, networks with multiple subsystems, and other complicated networked structures.

A multilayer network $M=(V_M,E_M,V,\mathtt{L})$ (see Fig.~\ref{fig1}(c)) has a set $V$ of nodes --- these are sometimes called ``physical nodes'', and each of them corresponds to an entity, such as a person --- that have instantiations as ``state nodes'' (i.e., node-layer tuples, which are elements of the set $V_M$) on layers in $\mathtt{L}$. One layer in the set $\mathtt{L}$ is a combination, through the Cartesian product $L_1 \times \dots \times L_d$, of elementary layers. The number $d$ indicates the number of types of layering; these are called ``aspects''.  
  A temporal network with one type of relationship has one type of layering, a time-independent network with multiple types of social relationships also has one type of layering, a multirelational network that changes in time has two types of layering, and so on. The set of state nodes
 in $M$ is $V_M \subseteq V \times L_1 \times \dots \times L_d$, and the set of 
 edges is $E_M \subseteq V_M \times V_M$. The edge $((i,\alpha),(j,\beta)) \in E_M$ indicates that there is an edge from node $j$ on layer $\beta$ to node $i$ on layer $\alpha$ (and vice versa, if $M$ is undirected). For example, in Fig.~\ref{fig1}(c), there is a directed intralayer edge from (A,1) to (B,1) and an undirected interlayer edge between (A,1) and (A,2). The multilayer network in Fig.~\ref{fig1}(c) has three layers, $|V| = 5$ physical nodes,
  $d = 1$ aspect, $|V_M| = 13$ state nodes, and $|E_M| = 20$ edges. To consider weighted edges, one proceeds as in ordinary graphs by defining a function $w: E_M \longrightarrow \mathbb{R}$. As in ordinary graphs, one can also incorporate self-edges and multi-edges.

Multilayer networks can include both intralayer edges (which have the same meaning as in graphs) and interlayer edges. The multilayer network in Fig.~\ref{fig1}(c) has $4$ directed intralayer edges, $10$ undirected intralayer edges, and $6$ undirected interlayer edges. In most studies thus far of multilayer representations of temporal networks, researchers have included interlayer edges only between state nodes in consecutive layers and only between state nodes that are associated with the same entity (see Fig.~\ref{fig1}(c)). However, this restriction is not always desirable (see \cite{valdano2015} for an example), and one can envision interlayer couplings that incorporate ideas like time horizons and interlayer edge weights that decay over time.
For convenience, many researchers have used undirected interlayer edges in multilayer analyses of temporal networks, but it is often desirable for such edges to be directed to reflect the arrow of time \cite{taylor2019}.
The sequence of network layers, which constitute time layers, can represent a discrete-time temporal network at different time instances or a continuous-time network in which one bins (i.e., aggregates) the network's edges to form a sequence of time windows with interactions in each window.

Each $d$-aspect multilayer network with the same number of nodes in each layer has an associated adjacency tensor $\mathcal{A}$ of order $2(d+1)$. For unweighted multilayer networks, each edge in $E_M$ is associated with a $1$ entry of $\mathcal{A}$, and the other entries (the ``missing'' edges) are $0$. If a multilayer network does not have the same number of nodes in each layer, one can add empty nodes so that it does, but the edges that are attached to such nodes are ``forbidden''. There has been some research on tensorial properties of $\mathcal{A}$ \cite{dedomenico2013mathematical} (and it is worthwhile to undertake further studies of them), but the most common approach for computations is to flatten $\mathcal{A}$ into a ``supra-adjacency matrix'' ${\bf A}_M$ \cite{kivela2014,whatis2018}, which is the adjacency matrix of the graph $G_M$ that is associated with $M$. The entries of diagonal blocks of ${\bf A}_M$ correspond to intralayer edges, and the entries of off-diagonal blocks correspond to interlayer edges. 


\subsubsection{Centrality, clustering, and large-scale network structures}

Following a long line of research in sociology \cite{kochen1978}, two important ingredients in the study of networks are examining (1) the importances (``centralities'') of nodes, edges, and other small network structures and the relationship of measures of importance to dynamical processes on networks and (2) the large-scale organization of networks \cite{newman2018book,faust1994}. 

Studying central nodes in networks is useful for numerous applications, such as ranking Web pages, football teams, or physicists \cite{gleich2015}. It can also help reveal the roles of nodes in networks, such as those that experience high traffic or help bridge different parts of a network \cite{newman2018book,faust1994}. Mesoscale features can impact network function and dynamics in important ways. Small subgraphs called ``motifs'' may appear frequently in some networks \cite{milo2002}, perhaps indicating fundamental structures such as feedback loops and other building blocks of global behavior \cite{golubitsky2}. Various types of larger-scale network structures, such as dense ``communities'' of nodes \cite{Porter2009, Fortunato2016} and core--periphery structures \cite{csermely2013,rombach2017}, are also sometimes related to dynamical modules (e.g., a set of synchronized neurons) or functional modules (e.g., a set of proteins that are important for a certain regulatory process) \cite{herbert1962}. A common way to study large-scale structures\footnote{There are recent theoretical advances on examining network structure amidst rich but noisy data \cite{newman2018}, and it is important for research on both network structure and dynamics to explicitly consider such scenarios.} 
 is inference using statistical models of random networks, such as through stochastic block models (SBMs) \cite{peixoto2017bayesian}. Much recent research has generalized the study of large-scale network structure to temporal and multilayer networks \cite{kivela2014,aleta2018,holme2019}. 

Various types of centrality --- including betweenness centrality \cite{tang2010,kim2012b}, Bonacich and Katz centrality \cite{lerman2010centrality,Grindrod_Higham_2014}, communicability \cite{Grindrod_Higham_2013}, PageRank \cite{walker2007ranking,rossi2012}, and eigenvector centrality \cite{praprotnik2015spectral,flores2018eigenvector} --- have been generalized to temporal networks using a variety of approaches. Such generalizations make it possible to examine how node importances change over time as network structure evolves.

In recent work, my collaborators and I used multilayer representations of temporal networks to generalize eigenvector-based centralities to temporal networks \cite{taylor2017,taylor2019}.\footnote{There is also much research on generalizing centralities (including eigenvector-based centralities \cite{taylor2019b}) to other types of multilayer networks, such as multiplex networks \cite{kivela2014,nutshell2019}.}
One computes the eigenvector-based centralities of nodes for a time-independent network as the entries of the ``dominant'' eigenvector, which is associated with the largest positive eigenvalue (by the Perron--Frobenius theorem, the eigenvalue with the largest magnitude is guaranteed to be positive in these situations) of a centrality matrix ${C}(\mathbf{A})$. Examples include eigenvector centrality (by using ${C}(\mathbf{A}) = \mathbf{A}$) \cite{bonacich1972}, hub and authority scores\footnote{Nodes that are good authorities tend to have good hubs that point to them, and nodes that are good hubs tend to point to good authorities.} (by using ${C}(\mathbf{A}) = \mathbf{A}\mathbf{A}^T$ for hubs and $\mathbf{A}^T\mathbf{A}$ for authorities) \cite{kleinberg1999}, and PageRank \cite{gleich2015}. 

Given a discrete-time temporal network in the form of a sequence of adjacency matrices ${\bf A}^{(t)} \in\mathbb{R}^{N\times N}$ for $t\in\{1,\dots,T\}$, where $a_{ij}^{(t)}$ denotes a directed edge from entity $i$ to entity $j$ in time layer $t$, we construct a ``supracentrality matrix'' $\mathbb{C}(\omega)$, which couples centrality matrices ${C}({\bf A}^{(t)})$ of the individual time layers. We then compute the dominant eigenvector of $\mathbb{C}(\omega)$, where $\omega$ is an interlayer coupling strength.\footnote{A major open problem in multilayer network analysis is the measurement and/or inference of values of $\omega$ (and generalizations of it in the form of coupling tensors) \cite{whatis2018}.} In \cite{taylor2017,taylor2019}, a key example was the ranking of doctoral programs in the mathematical sciences (using data from the Mathematics Genealogy Project \cite{mgp}), where an edge from one institution to another arises when someone with a Ph.D. from the first institution supervises a Ph.D. student at the second institution. By calculating time-dependent centralities, we can study how the rankings of mathematical-sciences doctoral programs change over time and the dependence of such rankings on the value of $\omega$. Larger values of $\omega$ impose more ranking consistency across time, so centrality trajectories are less volatile for larger $\omega$ \cite{taylor2017,taylor2019}.


Multilayer representations of temporal networks have been very insightful in the detection of communities and how they split, merge, and otherwise evolve over time. Numerous methods for community detection --- including inference via SBMs \cite{Peixoto2017}, maximization of objective functions (especially ``modularity'') \cite{mucha2010}, and methods based on random walks and bottlenecks to their traversal of a network \cite{multimap,jeub2017} --- have been generalized from graphs to multilayer networks. They have yielded insights in a diverse variety of applications, including brain networks \cite{muldoon2018}, granular materials \cite{lia2018}, political voting networks \cite{mucha-moody2013,mucha2010}, disease spreading \cite{marta2016}, and ecology and animal behavior \cite{pilosof2017,finn2019}. To assist with such applications, there are efforts to develop and analyze multilayer random-network models that incorporate rich and flexible structures \cite{bazzi2019}, such as diverse types of interlayer correlations.


\subsubsection{Activity-driven models} \label{act}

Activity-driven (AD) models of temporal networks \cite{perra2012} are a popular family of generative models that encode instantaneous time-dependent descriptions of network dynamics through a function called an ``activity potential'', which encodes the mechanism to generate connections and characterizes the interactions between entities in a network. An activity potential encapsulates all of the information about the temporal network dynamics of an AD model, making it tractable to study dynamical processes (such as ones from Section \ref{sec4}) on networks that are generated by such a model. It is also common to compare the properties of networks that are generated by AD models to those of empirical temporal networks \cite{holme2019}.

In the original AD model of Perra et al. \cite{perra2012}, one considers a network with $N$ entities, which we encode by the nodes. We suppose that node $i$ has an activity rate $a_i = \eta x_i$, which gives the probability per unit time to create new interactions with other nodes. The scaling factor $\eta$ ensures that the mean number of active nodes per unit time is $\eta\langle x\rangle N$, where $\langle x \rangle = \frac{1}{N}\sum_{i = 1}^N x_i$.
 We define the activity rates such that $x_i \in [\epsilon,1]$, where $\epsilon > 0$, and we assign each $x_i$ from a probability distribution $F(x)$ that can either take a desired functional form or be constructed from empirical data. The model uses the following generative process:
\begin{itemize}
\item{At each discrete time step (of length $\Delta t$), start with a network $G_t$ that consists of $N$ isolated nodes.
}
\item{With a probability $a_i\Delta t$ that is independent of other nodes, node $i$ is active and generates $m$ edges, each of which attaches to other nodes uniformly (i.e., with the same probability for each node) and independently at random (without replacement). Nodes that are not active can still receive edges from active nodes.
}
\item{At the next time step $t + \Delta t$, we delete all edges from $G_t$, so all interactions have a constant duration of $\Delta t$. We then generate new interactions from scratch. This is convenient, as it allows one to apply techniques from Markov chains.
}
\end{itemize}
Because entities in time step $t$ do not have any memory of previous time steps, $F(x)$ encodes the network structure and dynamics.

The AD model of Perra et al. \cite{perra2012} is overly simplistic, but it is amenable to analysis and has provided a foundation for 
many more general AD models, including ones that incorporate memory \cite{rizzo-siads2018}. In Section \ref{activity-simp}, I discuss a generalization of AD models to simplicial complexes \cite{petri2018} that allows one to study instantaneous interactions that involve three or more entities in a network.


\subsection{Continuous Time}

Many networked systems evolve continuously in time, but most investigations of time-dependent networks rely on discrete or discretized time. It is important to undertake more analysis of continuous-time temporal networks.

Researchers have examined continuous-time networks in a variety of scenarios. Examples include a compartmental model of biological contagions~\cite{Valdano2018}, a generalization of Katz centrality to continuous time \cite{Grindrod_Higham_2014}, generalizations of AD models (see Section \ref{act}) to continuous time \cite{zino2016,zino2017}, and rankings in competitive sports~\cite{Motegi2012}.

In a recent paper \cite{walid2018}, my collaborators and I formulated a notion of ``tie-decay networks'' for studying networks that evolve in
continuous time. They distinguished between \emph{interactions}, which they modeled as discrete contacts, and \emph{ties}, which encode relationships and their strength as a function of time. For example, perhaps the strength of a tie decays exponentially after the most recent interaction. More realistically, perhaps the decay rate depends on the weight of a tie, with strong ties decaying more slowly than weak ones. One can also use point-process models like Hawkes processes \cite{laub2015} to examine similar ideas using a node-centric perspective. 

Suppose that there are $N$ interacting entities, and let ${\bf B}(t)$ be the $N
\times N$ time-dependent, real, non-negative matrix whose entries
$b_{ij}(t)$ encode the tie strength between agents $i$ and $j$
at time $t$. In~\cite{walid2018}, we made the following simplifying assumptions:
\begin{enumerate}
\item{As in~\cite{Jin2001}, ties decay exponentially when there are no interactions: $\frac{{\mathrm d} b_{ij}}{{\mathrm d}t}=-\alpha
  b_{ij}$, where $\alpha \geq 0$ is the decay rate.} 
\item{If two entities interact at time $t=\tau$, the strength of the tie between them grows instantaneously by $1$.}
\end{enumerate}
See \cite{zuo2019} for a comparison of various choices, including those in~\cite{walid2018} and~\cite{Jin2001}, for tie evolution over time.

In practice (e.g., in data-driven applications), one obtains ${\bf B}(t)$ by discretizing time, so let's suppose that there is at
most one interaction during each time step of length $\Delta t$. This occurs, for example, in a Poisson process. Such time discretization is common in the
simulation of stochastic dynamical systems, such as in Gillespie
algorithms~\cite{Erban2007,porter2016,vester2015}. 
Consider an $N \times N$ matrix ${\bf A}(t)$ in which $a_{ij}(t)=1$ if node $i$ interacts with node $j$ at time $t$ and
$a_{ij}(t)=0$ otherwise. For a directed network, ${\bf A}(t)$ has exactly one nonzero entry during each time step when there is an interaction and no nonzero entries when there isn't one. For an undirected network, because of the symmetric nature of interactions, there are exactly two nonzero entries in time steps that include an interaction. We write
\begin{equation}
	  {\bf B}(\tdt)  = e^{-\alpha \Delta t}{\bf B}(t) + {\bf A}(\tdt)\,.
  \label{eq:Bupdate}
\end{equation}
Equivalently, if interactions between entities occur at times
$\tau^{\left(\ell\right)}$ such that $0\leq \tau^{\left(0\right)} < \tau^{\left(1\right)}
< \ldots < \tau^{\left(T\right)}$, then at time $t\geq
\tau^{\left(T\right)}$, we have
\begin{equation}   \label{eq:B_summation}
 	 {\bf B}(t) = \sum_{k = 0}^T e^{-\alpha (t - \tau^{\left(k\right)})} {\bf A}(\tau^{\left(k\right)})\,.
\end{equation}

In \cite{walid2018}, my coauthors and I generalized PageRank \cite{pagerank,gleich2015} to tie-decay networks. One nice feature of their tie-decay PageRank is that it is applicable not just to data sets, but also to data streams, as one updates the PageRank values as new data arrives. By contrast, one problematic feature of many methods that rely on multilayer representations of temporal networks is that one needs to recompute everything for an entire data set upon acquiring new data, rather than updating prior results in a computationally efficient way.


\section{Dynamical Processes on Networks} \label{sec4}

A dynamical process can be discrete, continuous, or some mixture of the two; it can also be either deterministic or stochastic. It can take the form of one or several coupled ordinary differential equations (ODEs), partial differential equations (PDEs), maps, stochastic differential equations, and so on. 

A dynamical process requires a rule for updating the states of its dependent variables with respect one or more independent variables (e.g., time), and one also has (one or a variety of) initial conditions and/or boundary conditions. To formalize a dynamical process on a network, one needs a rule for how to update the states of the nodes and/or edges. 

The nodes (of one or more types) of a network are connected to each other in nontrivial ways by one or more types of edges. This leads to a natural question: How does nontrivial connectivity between nodes affect dynamical processes on a network \cite{porter2016}? When studying a dynamical process on a network, the network structure encodes which entities (i.e., nodes) of a system interact with each other and which do not. If desired, one can ignore the network structure entirely and just write out a dynamical system. However, keeping track of network structure is often a very useful and insightful form of bookkeeping, which one can exploit to systematically explore how particular structures affect the dynamics of particular dynamical processes.

Prominent examples of dynamical processes on networks include coupled oscillators \cite{arenas-review,kuramoto-review2016}, games \cite{jackson2014}, and the spread of diseases \cite{kiss2017,pastor-satorras2015} and opinions \cite{sune-yy2018,loreto2009}. There is also a large body of research on the control of dynamical processes on networks \cite{barabasi-liu-review,motter-chaos}.

Most studies of dynamics on networks have focused on extending familiar models --- such as compartmental models of biological contagions \cite{kiss2017} or Kuramoto phase oscillators \cite{kuramoto-review2016} --- by coupling entities using various types of network structures, but it is also important to formulate new dynamical processes from scratch, rather than only studying more complicated generalizations of our favorite models. When trying to illuminate the effects of network structure on a dynamical process, it is often insightful to provide a baseline comparison by examining the process on a convenient ensemble of random networks \cite{porter2016}.


\subsection{An illustrative example: A threshold model of a social contagion}

A simple, but illustrative, dynamical process on a network is the Watts threshold model (WTM) of a social contagion \cite{sune-yy2018,porter2016}. It provides a framework for illustrating how network structure can affect state changes, such as the adoption of a product or a behavior, and for exploring which scenarios lead to ``virality'' (in the form of state changes of a large number of nodes in a network).

The original WTM \cite{watts2002}, a binary-state threshold model that resembles bootstrap percolation \cite{bootstrap}, has a deterministic update rule, so stochasticity can come only from other sources (see Section \ref{stoch}). In a binary state model, each node is in one of two states; see \cite{gleeson2013} for a tabulation of well-known binary-state dynamics on networks. The WTM is a modification of Mark Granovetter's threshold model for social influence in a fully-mixed population \cite{granovetter78}. See \cite{valente-book,kkt2003} for early work on threshold models on networks that developed independently from investigations of the WTM. Threshold contagion models have been developed for many scenarios, including contagions with multiple stages \cite{melnik2013}, models with adoption latency \cite{oh2018}, models with synergistic interactions \cite{juul2018}, and situations with hipsters (who may prefer to adopt a minority state) \cite{juul2019}. 

In a binary-state threshold model such as the WTM, each node $i$ has a threshold $R_i$ that one draws from some distribution. Suppose that $R_i$ is constant in time, although one can generalize it to be time-dependent. At any time, each node can be in one of two states: $0$ (which represents being inactive, not adopted, not infected, and so on) or $1$ (active, adopted, infected, and so on). A binary-state model is a drastic oversimplification of reality, but the WTM is able to capture two crucial features of social systems \cite{oliver1985}: interdependence (an entity's behavior depends on the behavior of other entities) and heterogeneity (as nodes with different threshold values behave differently). One can assign a seed number or seed fraction of nodes to the active state, and one can choose the initially active nodes either deterministically or randomly.
 
The states of the nodes change in time according to an update rule, which can either be synchronous (such that it is a map) or asynchronous (e.g., as a discretization of continuous time) \cite{porter2016}. In the WTM, the update rule is deterministic, so this choice affects only how long it takes to reach a steady state; it does not affect the steady state itself. With a stochastic update rule, the synchronous and asynchronous versions of ostensibly the ``same'' model can behave in drastically different ways \cite{melnik-sync}. In the WTM on an undirected network, to update the state of a node, one compares its fraction $s_i/k_i$ of active neighbors (where $s_i$ is the number of active neighbors and $k_i$ is the degree of node $i$) to the node's threshold $R_i$. An inactive node $i$ becomes active (i.e., it switches from state $0$ to state $1$) if $s_i/k_i \geq R_i$; otherwise, it stays inactive. The states of nodes in the WTM are monotonic, in the sense that a node that becomes active remains active forever. This feature is convenient for deriving accurate approximations for the global behavior of the WTM using branching-process approximations \cite{porter2016,gleeson2013} or when analyzing the behavior of the WTM using tools such as persistent homology \cite{taylor2015}.


\subsection{Stochastic processes} \label{stoch}

A dynamical process on a network can take the form of a stochastic process \cite{porter2016,newman2018book}. There are several possible sources of stochasticity: (1) choice of initial condition, (2) choice of which nodes or edges to update (when considering asynchronous updating), (3) the rule for updating nodes or edges, (4) the values of parameters in an update rule, and (5) selection of particular networks from a random-graph ensemble (i.e., a probability distribution on graphs). Some or all of these sources of randomness can be present when studying dynamical processes on networks. It is desirable to compare the sample mean of a stochastic process on a network to an ensemble average (i.e., to an expectation over a suitable probability distribution).

Prominent examples of stochastic processes on networks include percolation \cite{saberi2015}, random walks \cite{masuda2017}, compartment models of biological contagions \cite{kiss2017,pastor-satorras2015}, bounded-confidence models with continuous-valued opinions \cite{meng2018}, and other opinion and voter models \cite{porter2016,sune-yy2018,loreto2009,redner-review2018}.



\subsubsection{Example: A compartmental model of a biological contagion}

Compartmental models of biological contagions are a topic of intense interest in network science \cite{kiss2017,pastor-satorras2015,porter2016,newman2018book}. A compartment represents a possible state of a node; examples include susceptible, infected, zombified, vaccinated, and recovered. An update rule determines how a node changes its state from one compartment to another. One can formulate models with as many compartments as desired \cite{ccc}, but investigations of how network structure affects dynamics typically have employed examples with only two or three compartments \cite{kiss2017,pastor-satorras2015}. 

Researchers have studied various extensions of compartmental models, contagions on multilayer and temporal networks \cite{kivela2014,dedom2016,nutshell2019}, metapopulation models on networks \cite{colizza2007-meta} for simultaneously studying network connectivity and subpopulations with different characteristics, non-Markovian contagions on networks for exploring memory effects \cite{VanMieghem13}, and explicit incorporation of individuals with essential societal roles (e.g., health-care workers) \cite{scarpino2016}. As I discuss in Section \ref{multi}, one can also examine coupling between biological contagions and the spread of information (e.g., ``awareness'') \cite{bauch2015,funk-review2010}. One can also use compartmental models to study phenomena, such as dissemination of ideas on social media \cite{sharad2016} and forecasting of political elections \cite{volkening2019}, that are much different from the spread of diseases. 

One of the most prominent examples of a compartmental model is a susceptible--infected--recovered (SIR) model, which has three compartments. Susceptible nodes are healthy and can become infected, and infected nodes can eventually recover. The steady state of the basic SIR model on a network is related to a type of bond percolation \cite{grassberger83,kenah2007,trapman2007,lfd2013}. There are many variants of SIR models and other compartmental models on networks \cite{kiss2017}. See \cite{moritaSIS} for an illustration using susceptible--infected--susceptible (SIS) models. 

Suppose that an infection is transmitted from an infected node to a susceptible neighbor at a rate of $\lambda$. The probability of a transmission event on one edge between an infected node and a susceptible node in an infinitesimal time interval $\mathrm{d}t$ is $\lambda\, \mathrm{d}t$. Assuming that all infection events are independent, the probability that a susceptible node with $s$ infected neighbors becomes infected (i.e., for a node to transition from the $S$ compartment to the $I$ compartment, which represents both being infected and being infective) during $\mathrm{d}t$ is 
\begin{equation}
	1 - (1 - \lambda \, \mathrm{d}t)^s \to \lambda \, s \, \mathrm{d}t \quad \mathrm{as} \quad  \mathrm{d}t\to 0\,. \label{eqn1}
\end{equation}
If an infected node recovers at a constant rate of $\mu$, the probability that it switches from state $I$ to state $R$ in an infinitesimal time interval $\mathrm{d}t$ is $\mu \,\mathrm{d}t$.


\subsection{Deterministic dynamical systems}	

When there is no source of stochasticity, a dynamical process on a network is ``deterministic''. A deterministic dynamical system can take the form of a system of coupled maps, ODEs, PDEs, or something else. As with stochastic systems, the network structure encodes which entities of a system interact with each other and which do not. 

There are numerous interesting deterministic dynamical systems on networks --- just incorporate nontrivial connectivity between entities into your favorite deterministic model --- although it is worth noting that some stochastic features (e.g., choosing parameter values from a probability distribution or sampling choices of initial conditions) can arise in these models.


\subsubsection{Example: Coupled oscillators}

For concreteness, let's consider the popular setting of coupled oscillators. Each node in a network is associated with an oscillator, and we want to examine how network structure affects the collective behavior of the coupled oscillators. 
 
It is common to investigate various forms of synchronization (a type of coherent behavior), such that the rhythms of the oscillators adjust to match each other (or to match a subset of the oscillators) because of their interactions \cite{scholarsync}. A variety of methods, such as ``master stability functions'' \cite{pecora98}, have been developed to study the local stability of synchronized states and their generalizations \cite{arenas-review,porter2016}, such as cluster synchrony \cite{pecora-natcomms}. Cluster synchrony, which is related to work on ``coupled-cell networks'' \cite{golubitsky2}, uses ideas from computational group theory to find synchronized sets of oscillators that are not synchronized with other sets of synchronized oscillators. Many studies have also examined other types of states, such as ``chimera states'' \cite{pana2015}, in which some oscillators behave coherently but others behave incoherently. (Analogous phenomena sometimes occur in mathematics departments.)

A ubiquitous example is coupled Kuramoto oscillators on a network \cite{kuramoto-review2016,arenas2019,arenas-review}, which is perhaps the most common setting for exploring and developing new methods for studying coupled oscillators. (In principle, one can then build on these insights in studies of other oscillatory systems, such as in applications in neuroscience \cite{coombes2016}.) Coupled Kuramoto oscillators have been used for modeling numerous phenomena, including jetlag \cite{girvan-jetlag} and singing in frogs \cite{frogs2019}. Indeed, a ``Snowbird'' (SIAM) conference on applied dynamical systems would not be complete without at least several dozen talks on the Kuramoto model. In the Kuramoto model, each node $i$ has an associated phase $\theta_i(t) \in [0,2\pi)$.  
In the case of  ``diffusive'' coupling between the nodes\footnote{In this case, linearization yields Laplacian dynamics, which is closely related to a random walk on a network \cite{masuda2017}.}, the dynamics of the $i$th node is governed by the equation
\begin{align} \label{kuramoto-equ}
	\dot{\theta}_i := \frac{\mathrm{d}\theta_i}{\mathrm{d}t} = \omega_i + \sum_{j = 1}^N b_{ij} a_{ij} \sin(\theta_j - \theta_i)\,, \qquad i \in \{1,\ldots, N\}\,,
\end{align}
where one typically draws the natural frequency $\omega_i$ of node $i$ from some distribution $g(\omega)$, the scalar $a_{ij}$ is an adjacency-matrix entry of an unweighted network, $b_{ij}$ is the coupling strength on oscillator $i$ from oscillator $j$ (so $b_{ij}a_{ij}$ is an element of an adjacency matrix ${\bf W}$ of a weighted network), and $f_{ij}(y) = \sin(y)$ is the coupling function, which depends only on the phase difference between oscillators $i$ and $j$ because of the diffusive nature of the coupling. 

Once one knows the natural frequencies $\omega_i$, the model \eqref{kuramoto-equ} is a deterministic dynamical system, although there have been studies of coupled Kuramoto oscillators with additional stochastic terms \cite{gottwald2017}. Traditional studies of \eqref{kuramoto-equ} and its generalizations draw the natural frequencies from some distribution (e.g., a Gaussian or a compactly supported distribution), but some studies of so-called ``explosive synchronization'' (in which there is an abrupt phase transition from incoherent oscillators to synchronized oscillators) have employed deterministic natural frequencies \cite{arenas2019,bocca-kaboom}. The properties of the frequency distribution $g(\omega)$ have a significant effect on the dynamics of \eqref{kuramoto-equ}. Important features of $g(\omega)$ include whether it has compact support or not, whether it is symmetric or asymmetric, and whether it is unimodal or not \cite{kuramoto-review2016,strogatz2000}. 

The model \eqref{kuramoto-equ} has been generalized in numerous ways. For example, researchers have considered a large variety of coupling functions $f_{ij}$ (including ones that are not diffusive) and have incorporated an inertia term $\ddot{\theta}_i$ to yield a second-order Kuramoto oscillator at each node \cite{kuramoto-review2016}. The latter generalization is important for studies of coupled oscillators and synchronized dynamics in electric power grids \cite{timme-power}. Another noteworthy direction is the analysis of Kuramoto model on ``graphons'' (see, e.g., \cite{medvedev2014}), an important type of structure that arises in a suitable limit of large networks.


\subsection{Dynamical processes on multilayer networks} \label{multi}

An increasingly prominent topic in network analysis is the examination of how multilayer network structures --- multiple system components, multiple types of edges, co-occurrence and coupling of multiple dynamical processes, and so on --- affect qualitative and quantitative dynamics \cite{dedom2016,aleta2018,kivela2014}. For example, perhaps certain types of multilayer structures can induce unexpected instabilities or phase transitions in certain types of dynamical processes?

There are two categories of dynamical processes on multilayer networks: (1) a single process can occur on a multilayer network; or (2) processes on different layers of a multilayer network can interact with each other \cite{dedom2016}. An important example of the first category is a random walk, where the relative speeds and probabilities of steps within layers versus steps between layers affect the qualitative nature of the dynamics. This, in turn, affects methods (such as community detection \cite{jeub2017,multimap}) that are based on random walks, as well as anything else in which the diffusion is relevant \cite{dedomenico2014navigability,buldu-porter2018}. Two other examples of the first category are the spread of information on social media (for which there are multiple communication channels, such as Facebook and Twitter) and multimodal transportation systems \cite{barth2014}. For instance, a multilayer network structure can induce congestion even when a system without coupling between layers is decongested in each layer independently \cite{sole2016PRL}. Examples of the second category of dynamical process are interactions between multiple strains of a disease and interactions between the spread of disease and the spread of information \cite{bauch2015,funk-review2010,funk2015}. Many other examples have been studied \cite{aleta2018}, including coupling between oscillator dynamics on one layer and a biased random walk on another layer (as a model for neuronal oscillations coupled to blood flow) \cite{arenas-bold1}.

Numerous interesting phenomena can occur when dynamical systems, such as spreading processes, are coupled to each other \cite{bauch2015}. For example, the spreading of one disease can facilitate infection by another \cite{sanz2014dynamics}, and the spread of awareness about a disease can inhibit spread of the disease itself (e.g., if people stay home when they are sick) \cite{granell2013interplay}. Interacting spreading processes can also exhibit other fascinating dynamics, such as oscillations that are induced by multilayer network structures in a biological contagion with multiple modes of transmission \cite{tien2018} and novel types of phase transitions \cite{dedom2016}. 

A major simplification in most work thus far on dynamical processes on multilayer networks is a tendency to focus on toy models. For example, a typical study of coupled spreading processes may consider a standard (e.g., SIR) model on each layer, and it may draw the connectivity pattern of each layer from the same standard random-graph model (e.g., an Erd\H{o}s--R\'{e}nyi model or a configuration model). However, when studying dynamics on multilayer networks, it is particular important in future work to incorporate heterogeneity in network structure and/or dynamical processes. For instance, diseases spread offline but information spreads both offline and online, so investigations of coupled information and disease spread ought to consider fundamentally different types of network structures for the two processes.


\subsection{Metric graphs and waves on networks}

Network structures also affect the dynamics of PDEs on networks \cite{smilansky2006,herrmann2014,ide2014,malbor2014,miele2015}. Interesting examples include a study of a Burgers equation on graphs to investigate how network structure affects the propagation of shocks \cite{herrmann2014} and investigations of reaction--diffusion equations and Turing patterns on networks \cite{malbor2014,turing2015}. The latter studies exploit the rich theory of Laplacian dynamics on graphs (and concomitant ideas from spectral graph theory) \cite{masuda2017,piet-book} and examine the addition of nonlinear terms to Laplacians on various types of networks (including multilayer ones).

A mathematically oriented thread of research on PDEs on networks has built on ideas from so-called ``quantum graphs'' \cite{smilansky2006,kuchment2004} to study wave propagation on networks through the analysis of ``metric graphs''. Metric graphs differ from the usual ``combinatorial graphs'', which in other contexts are usually called simply ``graphs''.\footnote{Combinatorial graphs, and more general combinatorial objects, are my main focus in this chapter. This subsection is an exception.} In metric graphs, in addition to nodes and edges, each edge $e$ has a positive length $l_e \in (0,\infty]$. For many experimentally relevant scenarios (e.g., in models of circuits of quantum wires \cite{qwire-wiki}), there is a natural embedding into space, but metric graphs that are not embedded in space are also appropriate for some applications. 

As the nomenclature suggests, one can equip a metric graph with a natural metric. If a sequence $\{e_j\}_{j=1}^m$ of edges forms a path, the length of the path is $\sum_j l_j$. The distance $\rho(v_1,v_2)$ between two nodes, $v_1$ and $v_2$, is the minimum path length between them. We place coordinates along each edge, so we can compute a distance between points $x_1$ and $x_2$ on a metric graph even when those points are not located at nodes. Traditionally, one assumes that the infinite ends (which one can construe as ``leads'' at infinity, as in scattering theory) of infinite edges have degree $1$. It is also traditional to assume that there is always a positive distance between distinct nodes and that there are no finite-length paths with infinitely many edges. See \cite{kuchment2004} for further discussion.

To study waves on metric graphs, one needs to define operators, such as the negative second derivative or more general Schr\"odinger operators. This exploits the fact that there are coordinates for all points on the edges --- not only at the nodes themselves, as in combinatorial graphs. When studying waves on metric graphs, it is also necessary to impose boundary conditions at the nodes \cite{kuchment2004}.

Many studies of wave propagation on metric graphs have considered generalizations of nonlinear wave equations, such as the cubic nonlinear Schr\"odinger (NLS) equation \cite{nls-noja} and a nonlinear Dirac equation \cite{dirac2018}. The overwhelming majority of studies in metric graphs (with both linear and nonlinear waves) have focused on networks with a very small number of nodes, as even small networks yield very interesting dynamics. For example, Marzuola and Pelinovsky \cite{marzuola2017} analyzed symmetry-breaking and symmetry-preserving bifurcations of standing waves of the cubic NLS on a dumbbell graph (with two rings attached to a central line segment and Kirchhoff boundary conditions at the nodes). Kairzhan et al. \cite{goodman2019} studied the spectral stability of half-soliton standing waves of the cubic NLS equation on balanced star graphs. Sobirov et al. \cite{sobirov2016} studied scattering and transmission at nodes of sine--Gordon solitons on networks (e.g., on a star graph and a small tree).

A particularly interesting direction for future work is to study wave dynamics on large metric graphs. This will help extend investigations, as in ODEs and maps, of how network structures affect dynamics on networks to the realm of linear and nonlinear waves. One can readily formulate wave equations on large metric graphs by specifying relevant boundary conditions and rules at each junction. For example, Joly et al. \cite{joly2019} recently examined wave propagation of the standard linear wave equation on fractal trees. Because many natural real-life settings are spatially embedded (e.g., wave propagation in granular materials \cite{lia2018,leonard2014} and traffic-flow patterns in cities), it will be particularly valuable to examine wave dynamics on (both synthetic and empirical) spatially-embedded networks \cite{barthelemy2018}. Therefore, I anticipate that it will be very insightful to undertake studies of wave dynamics on networks such as random geometric graphs, random neighborhood graphs, and other spatial structures. A key question in network analysis is how different types of network structure affect different types of dynamical processes \cite{porter2016}, and the ability to take a limit as model synthetic networks become infinitely large (i.e., a thermodynamic limit) is crucial for obtaining many key theoretical insights. 


\section{Adaptive Networks} \label{sec5}

Dynamics of networks and dynamics on networks do not occur in isolation; instead, they are coupled to each other. Researchers have studied the coevolution of network structure and the states of nodes and/or edges in the context of ``adaptive networks'' (which are also known as ``coevolving networks'') \cite{gross2007adaptive,sayama2013modeling}. Whether it is sensible to study a dynamical process on a time-independent network, a temporal network with frozen (or no) node or edge states, or an adaptive network depends on the relative time scales of the dynamics of network structure and the states of nodes and/or edges of a network. See \cite{porter2016} for a brief discussion.

Models in the form of adaptive networks provide a promising mechanistic approach to simultaneously explain both structural features (e.g., degree distributions and temporal features (e.g., burstiness) of empirical data \cite{aoki2016}. Incorporating adaptation into conventional models can produce extremely interesting and rich dynamics, such as the spontaneous development of extreme states in opinion models \cite{sayama2015}.

Most studies of adaptive networks that include some analysis (i.e., that go beyond numerical computations) have employed rather artificial adaption rules for adding, removing, and rewiring edges. This is relevant for mathematical tractability, but it is important to go beyond these limitations by considering more realistic types of adaptation and coupling between network structure (including multilayer structures, as in \cite{berner2019}) and the states of nodes and edges. 


\subsection{Contagion models}\label{contag}

When people are sick, they stay home from work or school. People also form and remove social connections (both online and offline) based on observed opinions and behaviors. To study these ideas using adaptive networks, researchers have coupled models of biological and social contagions with time-dependent networks \cite{sune-yy2018,porter2016}.

An early example of an adaptive network of disease spreading is the susceptible--infected (SI) model in Gross et al. \cite{gross2006PRL}. In this model, susceptible nodes sometimes rewire their incident edges to ``protect themselves''. Suppose that we have an $N$-node network with a constant number of undirected edges. Each node is either susceptible (i.e., of type $S$) or infected (i.e., of type $I$). At each time step, and for each edge --- so-called ``discordant edges'' --- between nodes of different types, the susceptible node becomes infected with probability $\lambda$. For each discordant edge, with some probability $\kappa$, the incident susceptible node breaks the edge and rewires to some other susceptible node. This is a ``rewire-to-same'' mechanism, to use the language from some adaptive opinion models \cite{durrett2012graph,kureh2019}. (In this model, multi-edges and self-edges are not allowed.) During each time step, infected nodes can also recover to become susceptible again.

Gross et al. \cite{gross2006PRL} studied how the rewiring probability affects the ``basic reproductive number'', which measures how many secondary infections on average occur for each primary infection \cite{kiss2017,pastor-satorras2015,ccc}. This scalar quantity determines the size of a critical infection probability $\lambda_*$ to maintain a stable epidemic (as determined traditionally using linear stability analysis of an endemic state). A high rewiring rate can significantly increase $\lambda_*$ and thereby significantly reduce the prevalence of a contagion. Although results like these are perhaps intuitively clear, other studies of contagions on adaptive networks have yielded potentially actionable (and arguably nonintuitive) insights. For example, Scarpino et al. \cite{scarpino2016} demonstrated using an adaptive compartmental model (along with some empirical evidence) that the spread of a disease can accelerate when individuals with essential societal roles (e.g., health-care workers) become ill and are replaced with healthy individuals.


\subsection{Opinion models}  	

Another type of model with many interesting adaptive variants are opinion models \cite{loreto2009,porter2016}, especially in the form of generalizations of classical voter models \cite{redner-review2018}.

Voter dynamics were first considered in the 1970s by Clifford and Sudbury \cite{cliff1973} as a model for species competition, and the dynamical process that they introduced was dubbed ``the voter model"\footnote{There are several variants of ``the'' voter model, depending on choices such as whether one selects nodes or edges at random, that have substantively different qualitative dynamics \cite{porter2016,masuda2017}.} by Holley and Liggett shortly thereafter \cite{holley1975}. Voter dynamics are fun and are popular to study \cite{redner-review2018}, although it is questionable whether it is ever possible to genuinely construe voter models as models of voters \cite{ramasco2014}.

Holme and Newman \cite{petter2006} undertook an early study of a rewire-to-same adaptive voter model. Inspired by their research, Durrett et al. \cite{durrett2012graph} compared the dynamics from two different types of rewiring in an adaptive voter model. In each variant of their model, one considers an $N$-node network and supposes that each node is in one of two states. The network structure and the node states coevolve. Pick an edge uniformly at random. If this edge is discordant, then with probability $1 - \kappa$, one of its incident nodes adopts the opinion state of the other. Otherwise, with complementary probability $\kappa$, a rewiring action occurs: one removes the discordant edge, and one of the associated nodes attaches to a new node either through a rewire-to-same mechanism (choosing uniformly at random among the nodes with the same opinion state) or through a ``rewire-to-random'' mechanism (choosing uniformly at random among all nodes). As with the adaptive SI model in \cite{gross2006PRL}, self-edges and multi-edges are not allowed. 

The models in \cite{durrett2012graph} evolve until there are no discordant edges. There are several key questions. Does the system reach a consensus (in which all nodes are in the same state)? If so, how long does it take to converge to consensus? If not, how many opinion clusters (each of which is a connected component, perhaps interpretable as an ``echo chamber'', of the final network) are there at steady state? How long does it take to reach this state? The answers and analysis are subtle; they depend on the initial network topology, the initial conditions, and the specific choice of rewiring rule. As with other adaptive network models, researchers have developed some nonrigorous theory (e.g., using mean-field approximations and their generalizations) on adaptive voter models with simplistic rewiring schemes, but they have struggled to extend these ideas to models with more realistic rewiring schemes. There are very few mathematically rigorous results on adaptive voter models, although there do exist some, under various assumptions on initial network structure and edge density \cite{basu2017evolving}.

Researchers have generalized adaptive voter models to consider more than two opinion states \cite{shi2013} and more general types of rewiring schemes \cite{malik2016}. As with other adaptive networks, analyzing adaptive opinion models with increasingly diverse types of rewiring schemes (ideally with a move towards increasing realism) is particularly important. In \cite{kureh2019}, Yacoub Kureh and I studied a variant of a voter model with nonlinear rewiring (where the probability that a node rewires or adopts is a function of how well it ``fits in'' within its neighborhood), including a ``rewire-to-none'' scheme to model unfriending and unfollowing in online social networks. It is also important to study adaptive opinion models with more realistic types of opinion dynamics. A promising example is adaptive generalizations of bounded-confidence models (see the introduction of \cite{meng2018} for a brief review of bounded-confidence models), which have continuous opinion states, with nodes interacting either with nodes or with other entities (such as media \cite{brooks2019}) whose opinion is sufficiently close to theirs. A recent numerical study examined an adaptive bounded-confidence model \cite{brede2019}; this is an important direction for future investigations.


\subsection{Synchronization of adaptive oscillators} 

It is also interesting to examine how the adaptation of oscillators --- including their intrinsic frequencies and/or the network structure that couples them to each other --- affects the collective behavior (e.g., synchronization) of a network of oscillators \cite{kuramoto-review2016}. Such ideas are useful for exploring mechanistic models of learning in the brain (e.g., through adaptation of coupling between oscillators to produce a desired limit cycle \cite{stroud2018}).

One nice example is by Skardal et al.~\cite{skardal2014}, who examined an adaptive model of coupled Kuramoto oscillators as a toy model of learning. First, we write the Kuramoto system as
\begin{align} \label{kuramoto2}
	\frac{{\mathrm d}{\theta}_i}{{\mathrm d}t} = \omega_i + \sum_{j = 1}^N f_{ij}(\theta_j - \theta_i)\,, \qquad i \in \{1,\ldots, N\}\,,
\end{align}
where $f_{ij}$ is a $2\pi$-periodic function of the phase difference between oscillators $i$ and $j$. One way to incorporate adaptation is to define an ``order parameter'' $r_i$ (which, in its traditional form, quantifies the amount of coherence of the coupled Kuramoto oscillators \cite{kuramoto-review2016}) for the $i$th oscillator by
\begin{align*}
	r_i = \sum_{j = 1}^N b_{ij} a_{ij} e^{\mathrm{i}\theta_j}\,, \quad i \in \{1,\ldots, N\}
\end{align*}
and to consider the following dynamical system:
\begin{align} \label{hebb}
	\frac{{\mathrm d}{\theta}_i}{{\mathrm d}t} = \omega_i + \lambda_D^{-1} \,{\rm{Im}}(z_i e^{-\mathrm{i} \theta_i})  \,, \quad
	\tau \frac{{\mathrm d}{z}_i}{{\mathrm d}t} = r_i - z_i \,, \quad
	T \frac{{\mathrm d}{b}_{ij}}{{\mathrm d}t} = \alpha + \beta \,{\rm{Re}}(r_i z_i^*) - b_{ij}\,, 
\end{align}	
where $\rm{Re}(\zeta)$ denotes the real part of a quantity $\zeta$ and $\rm{Im}(\zeta)$ denotes its imaginary part. In the model \eqref{hebb}, $\lambda_D$ denotes the largest positive eigenvalue of the adjacency matrix ${\bf A}$, the variable $z_i(t)$ is a time-delayed version of $r_i$ with time parameter $\tau$ (with $\tau \rightarrow 0$ implying that $z_i \rightarrow r_i$), and $z_i^*$ denotes the complex conjugate of $z_i$. One draws the frequencies $\omega_i$ from some distribution (e.g., a Lorentz distribution, as in \cite{skardal2014}), and we recall that $b_{ij}$ is the coupling strength on oscillator $i$ from oscillator $j$. The parameter $T$ gives an adaptation time scale, and $\alpha \in \mathbb{R}$ and $\beta \in \mathbb{R}$ are parameters (which one can adjust to study bifurcations). Skardal et al. \cite{skardal2014} interpreted scenarios with $\beta > 0$ as ``Hebbian'' adaptation (see \cite{choe2014}) and scenarios with $\beta < 0$ as anti-Hebbian adaptation, as they observed that oscillator synchrony is promoted when $\beta > 0$ and inhibited when $\beta < 0$.


\section{Higher-Order Structures and Dynamics} \label{sec6}

Most studies of networks have focused on networks with pairwise connections, in which each edge (unless it is a self-edge, which connects a node to itself) connects exactly two nodes to each other. However, many interactions --- such as playing games, coauthoring papers and other forms of collaboration, and horse races --- often occur between three or more entities of a network. To examine such situations, researchers have increasingly studied ``higher-order'' structures in networks, as they can exert a major influence on dynamical processes.


\subsection{Hypergraphs}

Perhaps the simplest way to account for higher-order structures in networks is to generalize from graphs to ``hypergraphs'' \cite{newman2018book}. Hypergraphs possess ``hyperedges'' that encode a connection between on arbitrary number of nodes, such as between all coauthors of a paper. This allows one to make important distinctions, such as between a $k$-clique (in which there are pairwise connections between each pair of nodes in a set of $k$ nodes) and a hyperedge that connects all $k$ of those nodes to each other, without the need for any pairwise connections.

One way to study a hypergraph is as a ``bipartite network'', in which nodes of a given type can be adjacent only to nodes of another type. For example, a scientist can be adjacent to a paper that they have written \cite{newman2001}, and a legislator can be adjacent to a committee on which they sit \cite{porter2005}. It is important to generalize ideas from graph theory to hypergraphs, such as by developing models of random hypergraphs \cite{newman2009hyper,chodrow2019,chodrow2019b}.


\subsection{Simplicial complexes}

Another way to study higher-order structures in networks is to use ``simplicial complexes'' \cite{otter2017,giusti-review,ghrist-online-book}. A simplicial complex is a space that is built from a union of points, edges, triangles, tetrahedra, and higher-dimensional polytopes (see Fig.~\ref{fig1}d). Simplicial complexes approximate topological spaces and thereby capture some of their properties.

A \emph{$p$-dimensional simplex} (i.e., a \emph{$p$-simplex}) is a $p$-dimensional polytope that is the convex hull of its $p+1$ vertices (i.e., nodes). A \emph{simplicial complex} $K$ is a set of simplices such that (1) every face of a simplex from $S$ is also in $S$ and (2) the intersection of any two simplices $\sigma_1, \sigma_2 \in S$ is a face of both $\sigma_1$ and $\sigma_2$. An increasing sequence $K_1 \subset K_2\subset \cdots\subset K_l $ of simplicial complexes forms a \emph{filtered simplicial complex}; each $K_i$ is a \emph{subcomplex}. As discussed in \cite{otter2017} and references therein, one can examine the homology of each subcomplex. In studying the homology of a topological space, one computes topological invariants that quantify features of different dimensions \cite{ghrist-online-book}. One studies ``persistent homology'' (PH) of a filtered simplicial complex to quantify the topological structure of a data set (e.g., a point cloud) across multiple scales of such data. The goal of such ``topological data analysis'' (TDA) is to measure the ``shape'' of data in the form of connected components, ``holes'' of various dimensionality, and so on \cite{otter2017}. From the perspective of network analysis, this yields insight into types of large-scale structure that complement traditional ones (such as community structure). See \cite{topaz-dsweb} for a friendly, nontechnical introduction to TDA.

A natural goal is to generalize ideas from network analysis to simplicial complexes. Important efforts include generalizing configuration models of random graphs \cite{fosdick2018} to random simplicial complexes \cite{bob2018,courtney2016}; generalizing well-known network growth mechanisms, such as preferential attachment \cite{ginestra1}; and developing geometric notions, like curvature, for networks \cite{samal2018}. An important modeling issue when studying higher-order network data is the question of when it is more appropriate (or convenient) to use the formalisms of hypergraphs or simplicial complexes.

The computation of PH has yielded insights on a diverse set of models and applications in network science and complex systems. Examples include granular materials \cite{konst2014,lia2018}, functional brain networks \cite{giusti-review,sizemore-netneuro-review}, quantification of ``political islands'' in voting data \cite{feng2019}, percolation theory \cite{speidel2018}, contagion dynamics \cite{taylor2015}, swarming and collective behavior \cite{topaz2015}, chaotic flows in ODEs and PDEs \cite{ks2019}, diurnal cycles in tropical cyclones \cite{tym2019}, and mathematics education \cite{bassett2019}. See the introduction to \cite{otter2017} for pointers to numerous other applications.

Most uses of simplicial complexes in network science and complex systems have focused on TDA (especially the computation of PH) and its applications \cite{otter2017,petri-review,salnikov2018}. In this chapter, however, I focus instead on a somewhat different (and increasingly popular) topic: the generalization of dynamical processes on and of networks to simplicial complexes to study the effects of higher-order interactions on network dynamics. Simplicial structures influence the collective behavior of the dynamics of coupled entities on networks (e.g., they can lead to novel bifurcations and phase transitions), and they provide a natural approach to analyze $p$-entity interaction terms, including for $p \geq 3$, in dynamical systems. Existing work includes research on linear diffusion dynamics (in the form of Hodge Laplacians, such as in \cite{hodge2018}) and generalizations of a variety of other popular types of dynamical processes on networks.

	
\subsection{Coupled phase oscillators with $p$-body interactions with $p \geq 3$} 

Given the ubiquitous study of coupled Kuramoto oscillators \cite{kuramoto-review2016}, a sensible starting point for exploring the impact of simultaneous coupling of three or more oscillators on a system's qualitative dynamics is to study a generalized Kuramoto model. For example, to include both two-entity (``two-body'') and three-entity interactions in a model of coupled oscillators on networks, we write \cite{tanaka2011} 
\begin{equation} \label{three-body}
	\dot{{\bf x}}_i = {\bf f}_i({\bf x}_i) + \sum_{j,k} {\bf W}_{ijk}({\bf x}_i,{\bf x}_j,{\bf x}_k)\,,
\end{equation}
where ${\bf f}_i$ describes the dynamics of oscillator $i$ and the three-oscillator interaction term ${\bf W}_{ijk}$ includes two-oscillator interaction terms ${\bf W}_{ij}({\bf x}_i,{\bf x}_j)$ as a special case.

An example of $N$ coupled Kuramoto oscillators with three-term interactions is \cite{tanaka2011}
\begin{align}
	\dot{\theta}_i &= \omega_i + \frac{1}{N}\sum_j\left[a_{ij}\sin(\theta_{ji} + \alpha_{1ij}) + b_{ij}\sin(2\theta_{ji} + \alpha_{2ij})\right] \notag \\
		&\quad + \frac{1}{N^2}\sum_{j,k}\left[c_{ijk}\sin(\theta_{ji} + \alpha_{3ijk})\cos(\theta_{ki} + \alpha_{4ijk})\right]\,,
\end{align}
where we draw the coefficients $a_{ij}$, $b_{ij}$, $c_{ijk}$, $\alpha_{1ij}$, $\alpha_{2ij}$, $\alpha_{3ijk}$, $\alpha_{4ijk}$ from various probability distributions. Including three-body interactions leads to a large variety of intricate dynamics, and I anticipate that incorporating the formalism of simplicial complexes will be very helpful for categorizing the possible dynamics.

In the last few years, several other researchers have also studied Kuramoto models with three-body interactions \cite{pikovsky2013,pikovsky2015,skardal2019}. A recent study \cite{skardal2019}, for example, discovered a continuum of abrupt desynchronization transitions with no counterpart in abrupt synchronization transitions. There have been mathematical studies of coupled oscillators with interactions of three or more entities using methods such as normal-form theory \cite{bick2016} and coupled-cell networks \cite{golubitsky2}.

An important point, as one can see in the above discussion (which does not employ the mathematical formalism of simplicial complexes), is that one does not necessarily need to explicitly use the language of simplicial complexes to study interactions between three or more entities in dynamical systems. Nevertheless, I anticipate that explicitly incorporating the formalism of simplicial complexes will be useful both for studying coupled oscillators on networks and for other dynamical systems. In upcoming studies, it will be important to determine when this formalism helps illuminate the dynamics of multi-entity interactions in dynamical systems and when simpler approaches suffice.


\subsection{Social dynamics and simplicial complexes}  \label{activity-simp}

Several recent papers have generalized models of social dynamics by incorporating higher-order interactions \cite{petri2018,petri2019,horst2019,leonie2019}. For example, perhaps somebody's opinion is influenced by a group discussion of three or more people, so it is relevant to consider opinion updates that are based on higher-order interactions. Some of these papers use some of the terminology of simplicial complexes, but it is mostly unclear (except perhaps for \cite{horst2019}) how the models in them take advantage of the associated mathematical formalism, so arguably it often may be unnecessary to use such language. Nevertheless, these models are very interesting and provide promising avenues for further research.

Petri and Barrat \cite{petri2018} generalized activity-driven models to simplicial complexes. Such a simplicial activity-driven (SAD) model generates time-dependent simplicial complexes, on which it is desirable to study dynamical processes (see Section \ref{sec4}), such as opinion dynamics, social contagions, and biological contagions.

The simplest version of the SAD model is defined as follows.
\begin{itemize}
\item{Each node $i$ has an activity rate $a_i$ that we draw independently from a distribution $F(x)$.
}
\item{At each discrete time step (of length $\Delta t$), we start with $N$ isolated nodes. Each node $i$ is active with a probability of $a_i \Delta t$, independently of all other nodes. If it is active, it creates a $(p-1)$-simplex (forming, in network terms, a clique of $p$ nodes) 
with $p - 1$ other nodes that we choose uniformly and independently at random (without replacement). One can either use a fixed value of $p$ or draw $p$ from some probability distribution.
}
\item{At the next time step, we delete all edges, so all interactions have a constant duration. We then generate new interactions from scratch.
}
\end{itemize}
This version of the SAD model is Markovian, and it is desirable to generalize it in various ways (e.g., by incorporating memory or community structure).

Iacopini et al. \cite{petri2019} recently developed a simplicial contagion model that generalizes an SI process on graphs. Consider a simplicial complex $K$ with $N$ nodes, and associate each node $i$ with a state $x_i(t) \in \{0,1\}$ at time $t$. If $x_i(t) = 0$, node $i$ is part of the susceptible class $S$; if $x_i(t) = 1$, it is part of the infected class $I$. The density of infected nodes at time $t$ is $\rho(t) = \frac{1}{N}\sum_{i = 1}^N x_i(t)$. Suppose that there are $D$ parameters $\varpi_1, \ldots, \varpi_D$ (with $D \in \{1, \ldots, N-1\}$), where $\varpi_d$ represents the probability per unit time that a susceptible node $i$ that participates in a $d$-dimensional simplex $\sigma$ is infected from each of the faces of $\sigma$, under the condition that all of the other nodes of the face are infected. That is, $\varpi_1$ is the probability per unit time that node $i$ is infected by an adjacent node $j$ via the edge $(i,j)$. Similarly, $\varpi_2$ is the probability per unit time that node $i$ is infected via the 2-simplex $(i,j,k)$ in which both $j$ and $k$ are infected, and so on. The recovery dynamics, in which an infected node $i$ becomes susceptible again, proceeds as in the SIR model that I discussed in Section \ref{stoch}. One can envision numerous interesting generalizations of this model (e.g., ones that are inspired by ideas that have been investigated in contagion models on graphs).

	
\section{Outlook} \label{sec7}

The study of networks is one of the most exciting and rapidly expanding areas of mathematics, and it touches on myriad other disciplines in both its methodology and its applications. Network analysis is increasingly prominent in numerous fields of scholarship (both theoretical and applied), it interacts very closely with data science, and it is important for a wealth of applications.

My focus in this chapter has been a forward-looking presentation of ideas in network analysis. My choices of which ideas to discuss reflect their connections to dynamics and nonlinearity, although I have also mentioned a few other burgeoning areas of network analysis in passing. Through its exciting combination of graph theory, dynamical systems, statistical mechanics, probability, linear algebra, scientific computation, data analysis, and many other subjects --- and through a comparable diversity of applications across the sciences, engineering, and the humanities --- the mathematics and science of networks has plenty to offer researchers for many years.


\section*{Acknowledgements}

I thank Jes\'us Cuevas, Panos Kevrekidis, and Avadh Saxena for the invitation to write this book chapter. I acknowledge financial support from the National Science Foundation (grant number 1922952) through the Algorithms for Threat Detection (ATD) program. I thank Mariano Beguerisse D\'{i}az, Manlio De Domenico, James Gleeson, Petter Holme, Renaud Lambiotte, and Hiroki Sayama for their helpful comments. I am particularly grateful to Heather Brooks, Michelle Feng, Panos Kevrekidis, and Alice Schwarze for their thorough and insightful comments on drafts of this chapter.





\end{document}